# Robust anomalous Hall effect in ferromagnetic metal LiMn$_6$Sn$_6$ under high pressure

Lingling Gao[1#], Junwen Lai[2#], Dong Chen[3,4#], Cuiying Pei[1], Qi Wang[1,5], Yi Zhao[1], Changhua Li[1], Weizheng Cao[1], Juefei Wu[1], Yulin Chen[1,5,6], Xingqiu Chen[2,7], Yan Sun[2,7*], Claudia Felser[3*], and Yanpeng Qi[1,5,8*]

1. School of Physical Science and Technology, ShanghaiTech University, Shanghai 201210, China
2. Shenyang National Laboratory for Materials Science, Institute of Metal Research, Chinese Academy of Sciences, Shenyang, 110016, China
3. Max Planck Institute for Chemical Physics of Solids, Dresden 01187, Germany
4. College of Physics, Qingdao University, Qingdao 266071, China
5. ShanghaiTech Laboratory for Topological Physics, ShanghaiTech University, Shanghai 201210, China
6. Department of Physics, Clarendon Laboratory, University of Oxford, Parks Road, Oxford OX1 3PU, UK
7. School of Materials Science and Engineering, University of Science and Technology of China, Shenyang 110016, China
8. Shanghai Key Laboratory of High-resolution Electron Microscopy, ShanghaiTech University, Shanghai 201210, China

# These authors contributed to this work equally.
* Correspondence should be addressed to Y.P.Q. (qiyp@shanghaitech.edu.cn) or C.F. (claudia.felser@cpfs.mpg.de) or Y. S (sunyan@imr.ac.cn)

**ABSTRACT**

**Recently, the giant intrinsic anomalous Hall effect (AHE) has been observed in the materials with kagome lattice. In this study, we systematically investigate the influence of high pressure on the AHE in the ferromagnet LiMn$_6$Sn$_6$ with clean Mn kagome lattice. Our *in-situ* high-pressure Raman spectroscopy indicates that the crystal structure of LiMn$_6$Sn$_6$ maintains a hexagonal phase under high pressures up to 8.51 GPa. The anomalous Hall conductivity (AHC) $\sigma_{xy}^A$ remains around 150 $\Omega^{-1}$ cm$^{-1}$, dominated by the intrinsic mechanism. Combined with theoretical calculations, our results indicate that the stable AHE under pressure in LiMn$_6$Sn$_6$ originates from the robust electronic and magnetic structure.**

The anomalous Hall effect (AHE)[1, 2], arising from the "anomalous" transverse group velocity of carriers, has been experimentally observed in materials with broken time-reversal symmetry, typically in a ferromagnetic phase. Although it was experimentally discovered more than a century ago, the microscopic mechanism of AHE remains an unsolved topic in condensed matter physics[3, 4]. It has been generally accepted that spin–orbit coupling and spin splitting are two essential ingredients for the AHE. In general, the AHE can be broadly engendered by two classes of mechanisms[3]: the extrinsic disorder-induced effects (e.g., skew scattering and side jump)[5-7], or the intrinsic Berry-curvature effect[8]. Nevertheless, as the AHE continues to be discovered in various material systems, it not only deepens our understanding of such a striking electronic transport phenomenon[9], but also paves the way for the application of next-generation spintronic devices[10].

Recently, the giant intrinsic AHE induced by the large Berry curvature has been observed in the materials with kagome lattice[11], such as the bilayer Fe kagome ferromagnet $Fe_3Sn_2$,[12, 13] the noncollinear antiferromagnet $Mn_3Sn$[14] and the magnetic Weyl semimetal $Co_3Sn_2S_2$.[15-17] Among the kagome family, $RMn_6Sn_6$ (R = trivalent rare earth elements) has been a special one and attracted growing interest due to the pristine Mn kagome lattice[18-23]. We have successfully synthesized high-quality $RMn_6Sn_6$ (R = Tb, Dy, Ho) single crystals and observed large anomalous Hall conductivity (AHC) arising from the intrinsic mechanism[24]. More interestingly, the R elements in $RMn_6Sn_6$ family can be completely replaced by Li, Mg, or Ca,[19, 25, 26] which not only reduce the number of valence electrons but also change the magnetic states. Thus, the $RMn_6Sn_6$ family with clean Mn kagome lattice supply an excellent platform to tune electronic and magnetic states and explore larger AHE.

Pressure is a clean and useful means to tune the interatomic distance, engineer the electronic and, subsequently, the macroscopic physical properties of the system[27-30]. To our knowledge, only a few experimental observations of high pressure modulated AHE have been reported up to now[31-35]. In this paper, we focus on $LiMn_6Sn_6$, one member of $RMn_6Sn_6$ family with clean Mn kagome lattice, and study the pressure effect on the AHE through *in-situ* high-pressure Raman spectroscopy, Hall transport measurements and first-principles calculations. We find that the AHE is quite stable against external pressure within our measurement range, which has been shown to originate from the robust electronic and magnetic structure using density-functional theory (DFT). Our results demonstrate that $LiMn_6Sn_6$ with clean Mn kagome lattice displays excellent

AHE upon compression and potential applications to the next-generation spintronic devices.

The single crystals of LiMn$_6$Sn$_6$ were grown by the self-flux method and the details of crystal growth are illustrated in Ref. 19[19]. The single crystal diffraction patterns were obtained using a Bruker dual sources single crystal X-ray diffractometer at room temperature, and the X-ray source comes from a molybdenum target. An *in-situ* high-pressure Raman spectroscopy investigation was performed using a Raman spectrometer (Renishaw inVia, UK) with a laser excitation wavelength of 532 nm and low-wavenumber filter. A symmetric diamond anvil cell (DAC) with anvil culet sizes of 400 $\mu$m was used, with silicon oil as pressure transmitting medium. The high-pressure electrical transport measurements were performed on Quantum Design PPMS-9T. *In situ* high-pressure transport measurements were conducted on a nonmagnetic DAC with 600 $\mu$m-culet diamond. The schematic plot of DAC electrical transport measurement device can be found in Ref. 36.[36] A cubic BN/epoxy mixture layer was inserted between BeCu gaskets and Pt electrical leads as insulator layer. A freshly cleaved single-crystal piece of ~ 200×150×25 $\mu$m was loaded with NaCl powder as the pressure transmitting medium. A five-probe method was used to measure the longitudinal and Hall electrical resistivity. The longitudinal current applied within the *ab* plane (in-plane), and the magnetic field is along the *c* axis (out-of-plane). The pressure was determined by the ruby luminescence method[37]. In order to remove the longitudinal resistivity contribution due to voltage probe misalignment, we extracted the pure Hall resistivity by the equation $\rho_{yx} = [\rho(+\mu_0H) - \rho(-\mu_0H)]/2$. Correspondingly, the longitudinal resistivity component is obtained using $\rho_{xx}(\mu_0H) = [\rho(+\mu_0H) + \rho(-\mu_0H)]/2$.

To calculate the electronic band structure, we took the experimentally measured lattice constants as the starting point and relaxed the atomic positions. The pressure conditions were simulated by shrinking the volume of primitive cells with the relaxation of lattice constants and atomic positions. The electronic and magnetic structure calculations were performed by using the code of Vienna Ab initio Simulation Package (VASP)[38] based on density functional theory with projected augmented wave potential. The exchanged and correlation energies were considered in the generalized gradient approximation (GGA), following Perdew–Burke–Ernzerhof parametrization scheme[39]. The energy cut off of plane wave basis was set to be 500 eV. To calculate the intrinsic anomalous Hall effect, we projected the Bloch wave functions into maximally localized Wannier functions (MLWFs)[40]. The tight binding model Hamiltonians were constructed based on the overlap of MLWFs. Based tight binding model Hamiltonians, the intrinsic AHCs

calculated by the Kubo formula in linear response approximation[41]:

$$\sigma_{yx}(E_F) = e^2\hbar \left(\frac{1}{2\pi}\right)^3 \int_{\vec{k}} d\vec{k} \sum_{E(n,\vec{k})<E_F} f(n,\vec{k})\Omega_{n,yx}(\vec{k}) \qquad (1)$$

$$\Omega_{n,yx}^z(\vec{k}) = \text{Im} \sum_{n' \neq n} \frac{\langle u(n,\vec{k})|\hat{v}_y|u(n',\vec{k})\rangle\langle u(n',\vec{k})|\hat{v}_x|u(n,\vec{k})\rangle - (x \leftrightarrow y)}{\left(E(n,\vec{k})-E(n',\vec{k})\right)^2} \qquad (2)$$

where $\hat{v}_{x(y)} = \frac{1}{\hbar}\frac{\partial H(\vec{k})}{\partial k_{x(y)}}$ is the velocity operator, $E(n,\vec{k})$ is the eigenvalue for the n-th eigen states of $|u(n,\vec{k})\rangle$, and $f(n,\vec{k})$ is the Fermi-Dirac distribution. A dense k-grid of 250×250×250 was used in the integral.

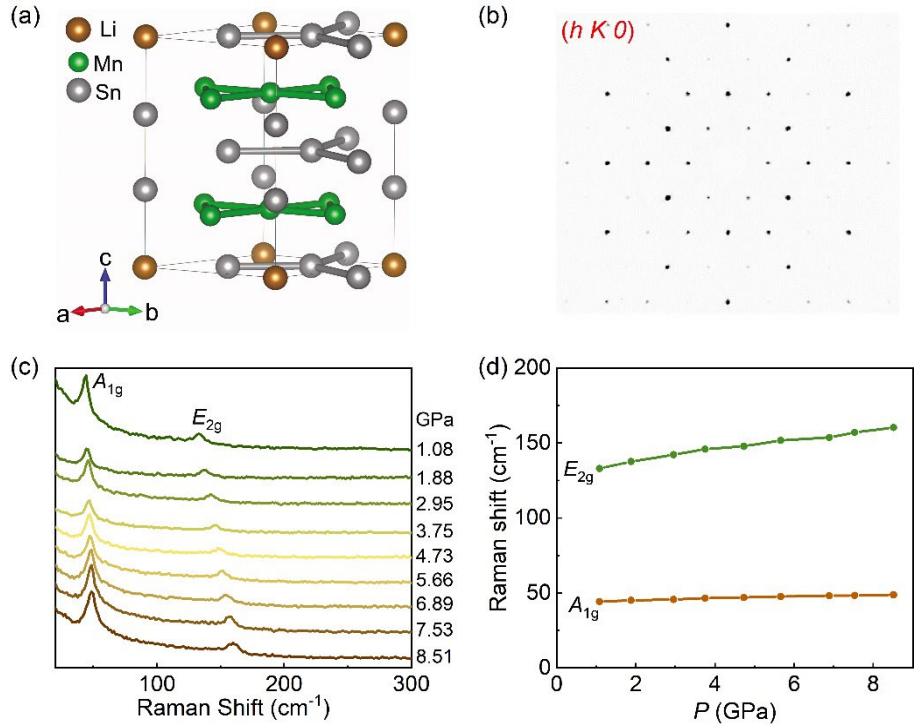

FIG. 1. (a) Crystal structure of LiMn$_6$Sn$_6$. The brown, green and silver balls represent Li, Mn and Sn, respectively. (b) Single crystal diffraction patterns in the reciprocal space along the (*h k 0*) direction. (c) Raman spectra at various pressures for LiMn$_6$Sn$_6$ at room temperature. (d) Raman shift for LiMn$_6$Sn$_6$ in compression; the vibration modes display in increasing wavenumber order.

As shown in Fig. 1(a), LiMn$_6$Sn$_6$ has the same crystal structure with the RMn$_6$Sn$_6$ compounds[24-26]. The single crystal diffraction pattern along the (*h k 0*) direction is displayed in Fig. 1(b), showing hexagonal symmetry. At ambient pressure, LiMn$_6$Sn$_6$ exhibits a ferromagnetic (FM) transition at $T_c$ = 380 K, and the easy plane is parallel to the *ab* plane[19]. The structure stability of LiMn$_6$Sn$_6$ is confirmed by *in-situ* Raman spectroscopy measurements. Fig. 1(c) shows the Raman spectra of LiMn$_6$Sn$_6$ under pressure up to 8.51 GPa. The assignments of the modes of LiMn$_6$Sn$_6$ at 1.08 GPa are given as $A_{1g}$ = 44.3 cm$^{-1}$ and $E_{2g}$ = 132.84 cm$^{-1}$. The profile of the spectra remains similar below 8.51 GPa, whereas the observed modes exhibit blue shift [Fig. 1(d)], thus showing the normal pressure behavior[42]. Our Raman results indicate that the structure of LiMn$_6$Sn$_6$ is robust and does not experience structural phase transition up to 8.51 GPa.

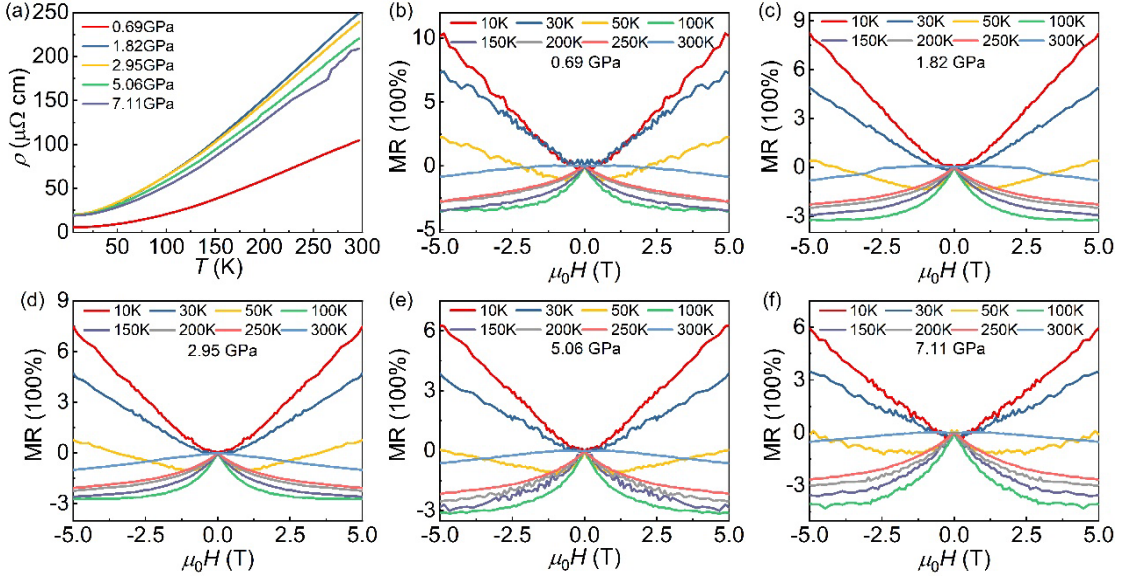

FIG. 2. (a) Temperature dependence of electric resistivity $\rho_{xx}(T)$ under high pressures. (b)-(f) Field dependence of magnetoresistance (MR) at various temperatures and selected pressures in Run-1.

We performed two independent runs of transport experiments (Run-1 and Run-2). Fig. 2(a) displays the longitudinal resistivity $\rho_{xx}(T)$ as a function of temperature from 1.8 K to 300 K in Run-1. All of the resistivity cuvers exhibit metallic behavior at selected pressures, which are consistent with resistivity curve at ambient pressure. The data of magnetoresistance (MR = [$\rho_{xx}(\mu_0H) - \rho_{xx}(0)$] / $\rho_{xx}(0) \times 100\%$) have been normalized in Figs. 2(b-f) and Fig. S1 (supplementary material) for Run-1 and Run-2, respectively. The MR curves also show similar characteristics to that under ambient pressure, and visibly change at 50 K for different pressures. When $T$ is below 50 K, the value of MR is positive, reveals the dominant role of Lorenz force in controlling MR process. At high temperature, it becomes negative owing to the suppressed spin scattering by magnetic field. The magnitude of the MR gradual decrease at high pressure up to 7.11 GPa.

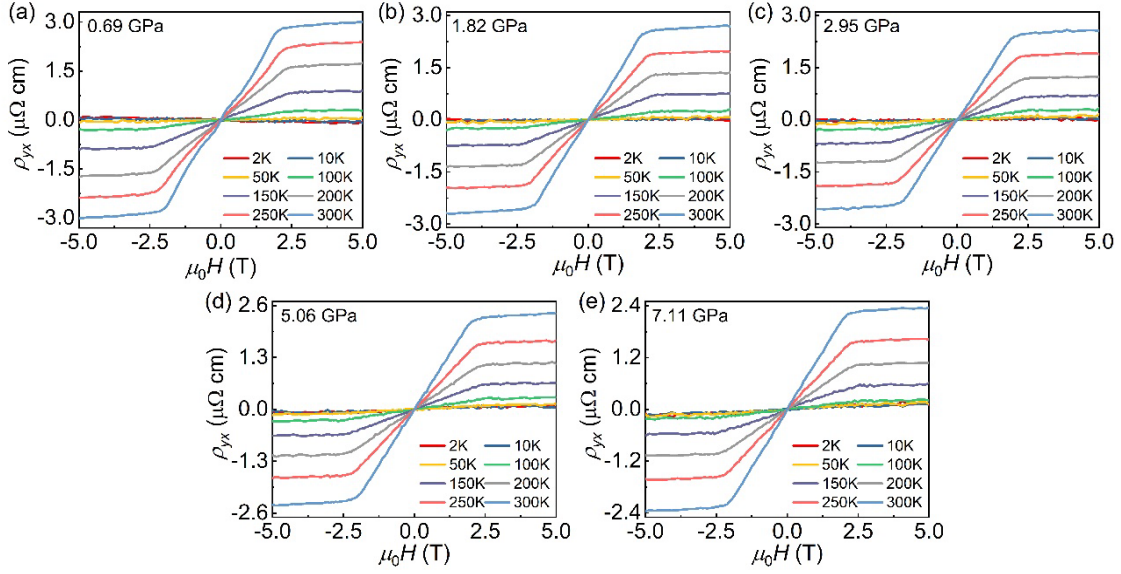

FIG. 3. (a)-(e) Field dependence of Hall resistivity $\rho_{yx}$ at various temperature and selected pressures in Run-1.

We performed the Hall effect measurements to evaluate the pressure effect on the AHE of LiMn$_6$Sn$_6$. Fig. 3 and Fig. S2 (supplementary material) show the field dependence of the Hall resistivity $\rho_{yx}(\mu_0 H)$ at selected pressures in Run-1 and Run-2, respectively. As can be seen, the $\rho_{yx}(\mu_0 H)$ data at each pressure share similar features as those at ambient pressure[19]. At high temperature, the $\rho_{yx}$ increases dramatically as the magnetic field increases and saturates at ~2 T, exhibiting typical features of AHE. Moreover, the saturated magnetic field $H_s$ of Hall resistivity is insensitive to pressure and remains almost unchanged up to 7.11 GPa, which illustrates the magnetization behavior of LiMn$_6$Sn$_6$ is stable against pressure. However, at low temperature, the anomalous Hall resistivity becomes too weak to be observed due to the reduction of resistivity, and the ordinary Hall resistivity is dominant. The slope of Hall resistivity is negative for 0.69 GPa and 0.41 GPa at 2 K, indicating an electron-type conduction in agreement with the ambient results[19]. With increasing the external pressure, the slope of the Hall resistance increases monotonically at 2 K and changes from negative to positive for 1.46 GPa, which implies a carrier-type inversion from electron- to hole-type. In addition, the saturation value of $\rho_{yx}$ at 300 K decreases slowly from 2.99 μΩ cm at 0.69 GPa to 2.35 μΩ cm at 7.11 GPa when the pressure increases.

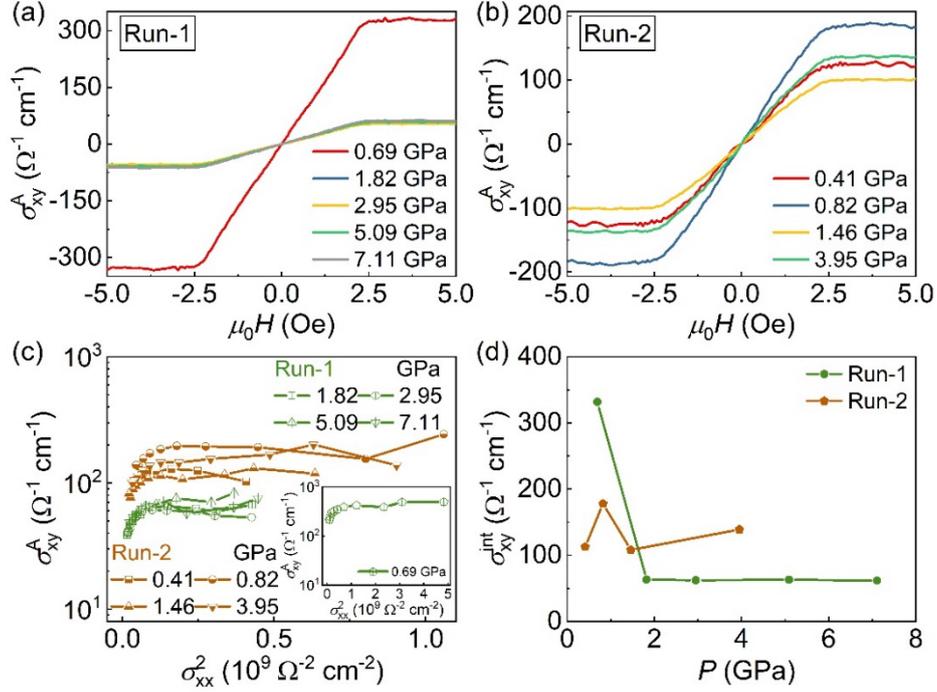

FIG. 4. Field-dependent anomalous Hall conductivity (AHC) $\sigma_{xy}^A$ at 200 K for selected pressures in Run-1 (a) and Run-2 (b). (c) Plot of AHC $\sigma_{xy}^A$ as a function of $\sigma_{xx}^2$. (d) Pressure-dependent intrinsic AHC $\sigma_{xy}^{int}$ in two runs.

The Hall conductivity can be obtained from $\sigma_{xy} = \rho_{yx} / (\rho_{yx}^2 + \rho_{xx}^2)$. Figs. 4(a) and 4(b) display the corresponding AHC $\sigma_{xy}^A$ at 200 K under different pressures in Run-1 and Run-2, respectively. In the Run-1, the saturation value of $\sigma_{xy}^A$ is 326 $\Omega^{-1}$ cm$^{-1}$ at 0.69 GPa, comparable with 380 $\Omega^{-1}$ cm$^{-1}$ at ambient pressure [19]. However, with increasing pressure, the saturation value rapid drop to about 65 $\Omega^{-1}$ cm$^{-1}$, and then remain almost unchanged. In the Run-2, the saturation value of $\sigma_{xy}^A$ first increases and then slowly decreases. This may be related to the incomplete contact between electrical leads and the sample at 0.41 GPa due to the use of solid NaCl as pressure transmitting medium. It is generally accepted that the total AHC consists of three terms in ferromagnetic conductor: $\sigma_H^A = \sigma_{int} + \sigma_{sk} + \sigma_{sj}$, where $\sigma_{int}$ is the intrinsic Karplus-Luttinger term, $\sigma_{sk}$ is the extrinsic skew scattering, and $\sigma_{sj}$ the generalized extrinsic side jump[3,4]. To separate the intrinsic from extrinsic contributions, we employ the so-called Tian-Ye-Jin (TYJ) scaling[4,13]: $\sigma_{xy}^A = f(\sigma_{xx,0})\sigma_{xx}^2 + \sigma_{xy}^{int}$, where $f(\sigma_{xx,0})$ is a function of the residual conductivity $\sigma_{xx,0}$, $\sigma_{xx}$ is the longitudinal conductivity and $\sigma_{xy}^{int}$ is the intrinsic AHC. We plot the $\sigma_{xy}^A$ as a function of $\sigma_{xx}^2$ for different pressures in Fig. 4(c). Because $\sigma_{xy}^{int}$ does not depend on the scattering rate, $\sigma_{xy}^{int}$ is then the remnant $\sigma_{xy}^A$ that is observed as $\sigma_{xx}^2 \to 0$. The $\sigma_{xy}^{int}$ as a function of pressure is displayed in Fig. 4(d). In the Run-1, the $\sigma_{xy}^{int}$ decreases rapidly from 330 $\Omega^{-1}$ cm$^{-1}$ at 0.69 GPa to ~65 $\Omega^{-1}$ cm$^{-1}$. With further compression, the $\sigma_{xy}^{int}$ keeps at ~65 $\Omega^{-1}$ cm$^{-1}$. In the Run-2, the $\sigma_{xy}^{int}$ basically keeps at ~150 $\Omega^{-1}$ cm$^{-1}$. As can be seen, the intrinsic AHC $\sigma_{xy}^{int}$ of LiMn$_6$Sn$_6$ is robust at high pressure.

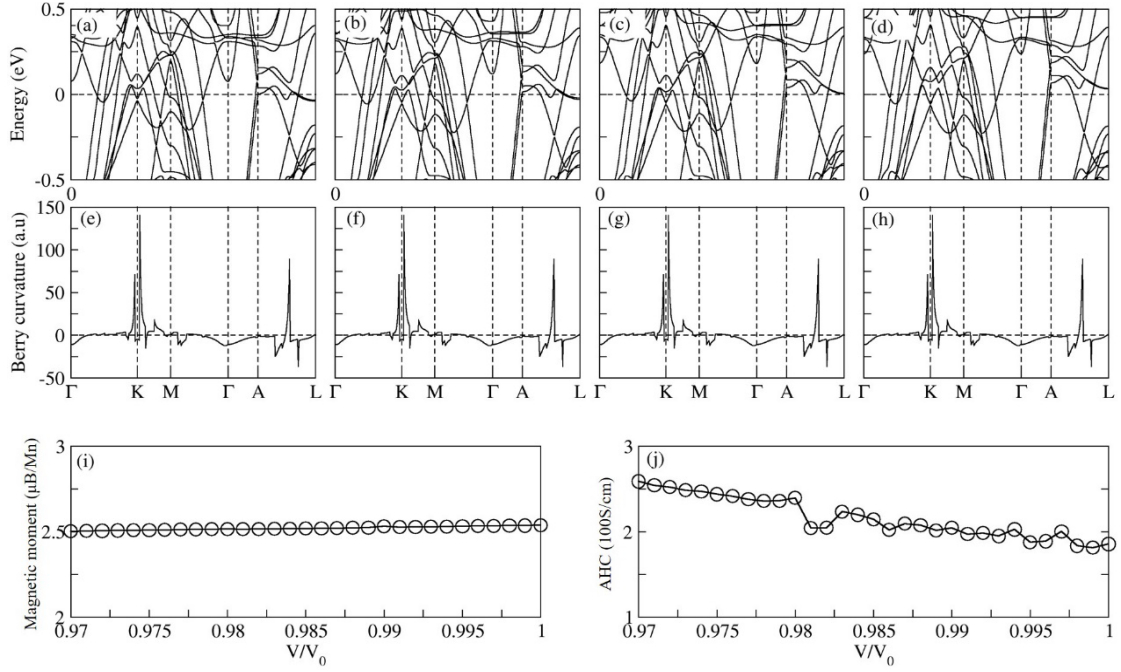

FIG. 5. Evolution of electronic structure, Berry curvature, magnetization, and AHC. (a-d) Energy dispersion along high symmetry lines for the cases of V/V$_0$=1.00, 0.99, 0.98, and 0.97, respectively. V$_0$ is the volume from experimental measurement at zero pressure. (e-h) Berry curvature ($\Omega_{yx}$ component) along high symmetry lines in the same condition with (a-d). (i, j) Volume-dependent magnetic moment and anomalous AHC, respectively.

To further understand the pressure effects, we performed the theoretical calculations. As the volume shrinks with pressure, there is almost no change for the electronic structure near the Fermi level, see the evolution of energy dispersion in Figs. 5(a-d). Correspondingly, the distribution of the $\Omega_{yx}$ component of Berry curvature is also robust, see Figs. 5(e-h). From the Berry curvature distribution, one can see that the three peaks of $\Omega_{yx}$ locate near K points and between A-L, which mainly originate from anti-crossings with tiny effective masses around the Fermi level. Since all three peaks are positive, the integral of Berry curvature in the whole k-space gives a positive AHC around 180 S/cm at zero pressure, in good agreement with that from Hall measurement. In addition, the calculated magnetic moment of LiMn$_6$Sn$_6$ is around 2.54 $\mu_B$/Mn, also close to the saturated magnetic moment (2.4 $\mu_B$/Mn) at ambient pressure[19]. Consistent with the results of volume-dependent electronic band structure and Berry curvature distributions, both magnetization and AHC are robust against pressure. As presented in Figs. 5(i-j), the magnetic moment and AHC are limited in the range of ~2.50 to ~2.54 $\mu_B$/Mn, and ~180 to 250 S/cm, respectively. Therefore, the stable AHE under perturbation of pressure in LiMn$_6$Sn$_6$ originates from the robust electronic and magnetic structure.

In conclusion, we study the pressure effect on the AHE of LiMn$_6$Sn$_6$ through *in-situ* high-pressure Raman spectroscopy, Hall transport measurements and first-principles calculations. The crystal structure and AHE of LiMn$_6$Sn$_6$ are robust under high pressure. According to the first-principles calculations, the stable AHE in LiMn$_6$Sn$_6$ originates from the robust electronic and magnetic structure under high pressure. This result shows that the AHC of LiMn$_6$Sn$_6$ is very stable under high pressure, which lays a good foundation for the development of spintronic devices in extreme environment.

See the supplementary material for detailed data of magnetoresistance and Hall resistivity as a function of magnetic field at different pressures and temperatures in Run-2.

This work was supported by the National Natural Science Foundation of China (Grants No. 52272265, 11974246, 12004252, U1932217), the National Key R&D Program of China (Grant No. 2018YFA0704300), Shanghai Science and Technology Plan (Grant No. 21DZ2260400) and Double First-Class Initiative Fund of ShanghaiTech University. Y.S. thanks the support from National Natural Science Foundation of China (Grants No. 52271016, 52188101). The authors thank the support from Analytical Instrumentation Center (# SPST-AIC10112914), SPST, ShanghaiTech University. C.F. thanks the European Research Council (ERC Advanced Grant No. 742068 'TOPMAT'), the DFG through SFB 1143 (project ID. 247310070), and the Würzburg-Dresden Cluster of Excellence on Complexity and Topology in Quantum Matter ct.qmat (EXC2147, project ID. 390858490).

## AUTHOR DECLARATIONS

**Conflict of Interest**

The authors have no conflicts to disclose.

## DATA AVAILABILITY

The data that support the findings of this study are available from the corresponding authors upon reasonable request.